\documentclass[conference]{IEEEtran}
\IEEEoverridecommandlockouts
\usepackage{cite}
\usepackage{amsmath,amssymb,amsfonts}
\usepackage{algorithmic}
\usepackage{graphicx}
\usepackage{textcomp}
\usepackage{xcolor}
\usepackage{relsize}
\usepackage{multirow}
\usepackage{makecell}
\usepackage{hyperref}
\usepackage{enumitem}
\usepackage{amsmath} 

\def\BibTeX{{\rm B\kern-.05em{\sc i\kern-.025em b}\kern-.08em
    T\kern-.1667em\lower.7ex\hbox{E}\kern-.125emX}}
\begin{document}

\title{\textit{AMuSE}: \textit{A}daptive \textit{Mu}ltimodal Analysis for \textit{S}peaker \textit{E}motion Recognition in Group Conversations}

% # Previous format for authors
% \author{\IEEEauthorblockN{Sidharth Anand}
%  email address
%  }
% \and
% \IEEEauthorblockN{Naresh Kumar Devulapally}
% \and
% \IEEEauthorblockN{Sreyasee Das Bhattacharjee}
% \and
% \IEEEauthorblockN{Junsong Yuan}

% New format 
\author{\IEEEauthorblockN{Naresh Kumar Devulapally, Sidharth Anand, Sreyasee Das Bhattacharjee, Junsong Yuan, Yu-Ping Chang}
%\IEEEauthorblockA{{Department of Computer Science and Engineering}}
\IEEEauthorblockA{The State University of New York at Buffalo.}}

% \IEEEauthorblockA{{{test}@buffalo.edu}}
\maketitle

\begin{abstract}

Analyzing individual emotions during group conversation is crucial in developing intelligent agents capable of natural human-machine interaction. While reliable emotion recognition techniques depend on different modalities (text, audio, video), the inherent heterogeneity between these modalities and the dynamic cross-modal interactions influenced by an individual's unique behavioral patterns make the task of emotion recognition very challenging. This difficulty is compounded in group settings, where the emotion and its temporal evolution are not only influenced by the individual but also by external contexts like audience reaction and context of the ongoing conversation. To meet this challenge, we propose a \textit{Multimodal Attention Network} (\textit{MAN}) that captures cross-modal interactions at various levels of spatial abstraction by jointly learning its interactive bunch of mode-specific \textit{Peripheral} and \textit{Central} networks. The proposed \textit{MAN} ``injects" cross-modal attention via its \textit{Peripheral} key-value pairs within each layer of a mode-specific \textit{Central} query network. The resulting cross-attended mode-specific descriptors are then combined using an \textit{Adaptive Fusion} (\textit{AF}) technique that enables the model to integrate the discriminative and complementary mode-specific data patterns within an instance-specific multimodal descriptor. Given a dialogue represented by a sequence of utterances, the proposed \textit{AMuSE} (\textit{A}daptive \textit{Mu}ltimodal Analysis for \textit{S}peaker \textit{E}motion) model condenses both spatial (within-mode and within-utterance) and temporal (across-mode and across-utterances in the sequence) features into two dense descriptors: speaker-level and utterance-level. This helps not only in delivering better classification performance ($3-5\%$ improvement in Weighted-F1 and $5-7\%$ improvement in Accuracy) in large-scale public datasets (MELD and IEMOCAP) but also helps the users in understanding the reasoning behind each emotion prediction made by the model via its \textit{Multimodal Explainability Visualization} module.
\end{abstract}


\begin{figure*}[t]
\centerline{\includegraphics[width=0.75\textwidth]{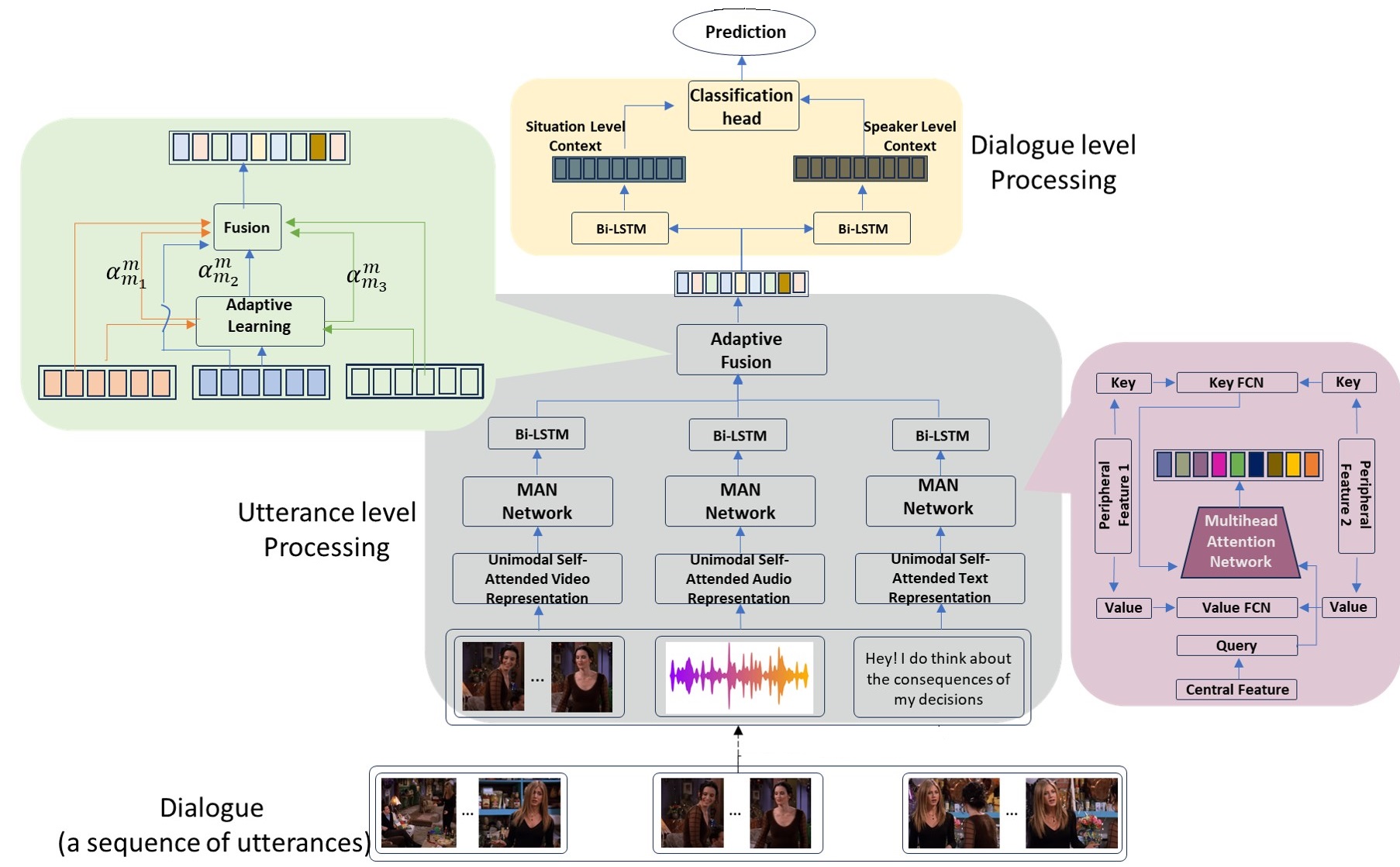}}
\caption{\small{Proposed \textit{AMuSE} Architecture that captures cross-modal interactions and their spatio-temporal evolution to predict the speaker's emotion in a conversation. }}
\label{fig:arch}
\end{figure*}

\section{Introduction}
Understanding the emotional nuances within conversations has become a pivotal task, with applications ranging from sentiment analysis in social media to affect-aware human-robot interactions.The complexities inherent in multi-party conversations, involving interactions between multiple speakers via various modalities like text, video, and audio, pose significant challenges for accurate emotion recognition. Extracting the subtle interplay of emotions from these diverse modalities necessitates a comprehensive approach that can effectively capture the spatio-temporal evolution of heterogeneous co-occurring mode-specific patterns and their mutual interactions across various modes, while also taking into consideration the unique and dynamic nature of the speaker's expressions.

It is particularly to note that the affective state of each individual evolves continuously during conversation. Such transitions depend on various intra (e.g. personal background, unique behavioral traits, and habits) and inter (e.g. audience behavior, their interpreting conducts) personal contexts as well as other environmental circumstances. While each speaker's visual and auditory cues offer valuable insights into their emotional states, the relationships between these modalities can be subtle and context-dependent. Additionally, the importance of different modes is not constant but varies in an instance-specific manner, depending on situations and contexts. For example, there may be a scenario in which a speaker's facial expression may be less explicitly representative of their emotion and its temporal evolution compared to their acoustic signal. In another scenario, a speaker may feel more restrained in expressing their emotions due to the conventional setting, such as at public gathering. In such cases, the surrounding contexts may provide important cues to facilitate accurate inferences regarding the speaker's emotion.  Thus the challenges related to \textit{the presence of strong heterogeneity among such cross-modal representations} and \textit{the influence of an individual's intra-personal and other situational contexts toward learning a variety of the cross-modal interaction patterns} are crucial yet under-studied.

%Thus the challenges due to \textit{the influence of an individual's intra-personal and other situational contexts toward learning a variety of the cross-modal interaction patterns} combined with \textit{the presence of strong heterogeneity among their mode-specific representations} are less studied for this task.

To address these challenges, we propose a two-level information integration technique. First, a Multimodal Attention Network (\textit{MAN}) is trained to bridge the heterogeneity gaps among the mode-specific representations by capturing the cross-modal interactions at
various levels of spatial abstraction. \textit{MAN} incorporates an interactive bunch of mode-specific \textit{Peripheral} and \textit{Central} networks, where the utterance-level mode-specific emotion patterns at every layer of a \textit{Central} network are attended by injected feedback from a \textit{Peripheral} network. This enables each mode-specific \textit{Central} network to prioritize the mode-invariant spatial (within utterance) details of the emotion patterns, while also retaining its mode-exclusive aspects within the learned model. Intuitively, this is necessary because while representing a specific emotional state, various modalities like text, audio, and video exhibit correlations that may be apparent at different levels of abstraction. While these correlating patterns are important, certain mode-exclusive cues (e.g., the non-verbal response of the audience) may also convey useful insights about the speaker's evolving emotional state at the next time-stamp. Second, an Adaptive Fusion (\textit{AF}) technique is employed, recognizing that not all modalities contribute equally to the process of emotion recognition for every query instance. The resulting cross-attended uni-modal feature descriptors derived from the mode-specific \textit{Central} networks are then interpolated via \textit{AF} in an instance-specific manner. These interpolated descriptors from the \textit{AF} are analyzed on both the conversational level and on a speaker level, allowing us to track emotional changes for the entire group and for each individual speaker.Thus, the  key contributions of the proposed model for the \textit{A}daptive \textit{Mu}ltimodal Analysis for \textit{S}peaker \textit{E}motion (aka \textit{AMuSE}), are: 
% \vspace{-2mm}
\begin{enumerate}[leftmargin=0.2cm,itemindent=0cm,labelwidth=\itemindent,align=left]
    \item \textbf{Cross-Attended Feature Representation via Multimodal Attention Network} that models cross-modal interaction by ``injecting" features from multiple \textit{Peripheral} networks into the layers of a \textit{Central} network. This helps the model to prioritize the mode-invariant spatial (within utterance) details of the emotion patterns, while also retaining its mode-exclusive aspects at various levels of abstraction.

\item \textbf{Adaptive Fusion (AF)} that interpolates the cross-attended mode-specific descriptors to combine the novel instance-specific and category-specific utterance-level spatial patterns within the learned multimodal descriptor.

    \item \textbf{Extensive Evaluations with Explainable Visualization} using publicly available (MELD \cite{poria2019meld}, IEMOCAP \cite{Busso2008}) datasets not only demonstrate an impressive classification performance ($3-5\%$ improvement in Weighted-F1 and $5-7\%$ improvement in Accuracy) of \textit{AMuSE}, its user-friendly \textit{interface} also facilitate the multimodal reasonings behind a specific prediction made by the model to deliver improved reliability on its decision.
\end{enumerate}

% spatio-temporalat various levels

\section{Related work}
Traditionally, works on Emotion Recognition in Conversations (ERC) have focused heavily on unimodal techniques, primarily due to the strength of natural language transcriptions or descriptors as a strong emotional indicator \cite{dialogrnn,cosmic,DBLP:conf/naacl/DevlinCLT19}. However, while text serves the purpose in simple scenarios, it often struggles to evaluate more complex human responses involving sarcasm or confusion, where more information can be gleaned about the speaker's emotional state by studying the tone, face, posture, and gestures \cite{ijcai2019p752}. Recent works \cite{chudasama2022m2fnet,li-etal-2022-emocaps} demonstrate the superiority of multimodal techniques. Most current ERC research has focused on modeling cross-modal interactions by either concatenating the processed unimodal feature vectors \cite{dialogrnn,dcrn,7837868} or by a predefined fixed combination (e.g., a weighted average of feature vectors) \cite{5674019,10.1145/2993148.2993176}. While better than unimodal approaches, these techniques ignore various levels of information that may be critical for the comprehensive modeling of spatial and temporal multimodal relationships. 

Though promising, many existing methods \cite{kiela2018efficient,shi2022learning} suffer from the weakness of an insufficient fusion of cross-modal interactions, unlike popular beliefs, which may not be uniform across instances or categories \cite{donohue2013cross} and vary given an individual's unique socio-behavioral responses. Furthermore, the precision vs. explainability trade-off continues to pose challenges for the systems. Toward addressing these, we leverage multimodal information from each sample to track the evolution of emotion over the conversation both for individual speakers and the group as a whole, while simultaneously providing meaningful insights that attempt to explain the model's decision.

\section{Proposed Method}
\textbf{Problem Definition:} Given a multi-party conversation represented as a sequence of utterances $\{u_j\} j\in \mathcal D$, the objective is to evaluate the dominant emotional state of the speaker for each utterance $u_j$. For brevity, we will henceforth omit the suffix $j$, and an arbitrary utterance $u_j$ will be represented as $u$ unless the suffix is specifically required. Each $u \in \mathcal{D}$ contains $(T, V, A)$, where $T$ is the \textit{text transcription}, $V$ is the \textit{video} and $A$ is the \textit{audio}.
%It is important to note that each dialog $d$, contains two temporal dimensions - a sequence of utterances $u_j$ and the temporal evolution of $T, V$, and $A$ within each utterance $u_j$.

\subsection{Unimodal Self-Attended Feature Representation}
\label{Uni_mode}
To capture the spatial evolution of information within each utterance, we propose a mode-specific unimodal feature representation scheme as described below

\subsubsection{Text Representation}
%To derive a representation $F^i_T$ of text $T$, represented as a series of $p$ words i.e., $T=\{\omega_1, \omega_2, ..., \omega_p\}$, we employ the pretrained model \cite{mpnet} to obtain the contextualized embedding for each word \textcolor{red}{suddenly f^T from $w_t   $} $f^T = \{f_1, f_2, ..., f_p\}, \forall f_i\in \mathbb R^{d_{T}}$.
To derive a compact descriptor for the text component $T$ represented as a sequence of $p$ words, i.e. $T=\{\omega_1, \omega_2, ..., \omega_p\}$, we employ the pretrained model \cite{mpnet} to obtain the fixed language embedding  $\mathbf f^{T}\in \mathbb R^{p\times d_{t}}$ for the text component $T$. The masked and permuted language modeling (MPNet) inherits the advantages of BERT \cite{devlin2018bert} and XLNet \cite{yang2019xlnet} by leveraging the dependency within the predicted tokens through permuted language modeling and utilizes the auxiliary position information to mitigate the position discrepancy. In fact, to explicitly capture the contextual meaning of each word in an utterance, the initial MPNet-based word embeddings are used as input to a Bi-directional-Long Short-Term Memory (Bi-LSTM) followed by the embedding layer to produce a derived word representation vector $\mathbf h_i$ for each $\omega_i$, which in turn develops a derived text representation vector $\mathbf w_0=[\mathbf h_1,\mathbf h_2, ...\mathbf h_p]$ for the text component $T$. Toward attaining an attention-aware text descriptor,  $\mathbf w_0$ is further processed through a M-layered attention  $F^{SA}_t$ network computed as $\mathbf w_{M}=F^{SA}_t(\mathbf w_{0},M)\in \mathbb R^{p\times d_t}$. An intermediate $m^{th}$ layer output in $F^{SA}_t$  is computed as $\mathbf w_{m+1}=\mathrm{linear}\bigg(\mathrm{softmax}\big(\frac{\mathbf w_m \mathbf w^T_m}{\sqrt{d_t}}\big)\mathbf w_m\bigg)$ and the resulting attention-enhanced average pooled text descriptor is defined as $\mathbf f^{T}=\mathbf w_{M}$.

\subsubsection{Video Representation}
 For the visual component $V$ of each utterance $u\in \mathcal D$, FFmpeg is used to identify $n$ keyframes and MTCNN \cite{zhang2016joint} is applied to extract the aligned faces from each keyframe. To represent the facial expression information within the context of the individuals' environment, each keyframe is then decomposed into two components: ``face frames", which is a derived frame containing only the face regions of the keyframes; "back frames", which captures the background environment by removing all identified faces. JAA-Net \cite{jaanet}, which jointly performs Action Units (AU) detection and facial landmark detection, is employed to extract AUs from each of these ``face frames". Thus, the visual content $V$ of $u$ is represented in terms of two equal-sized derived frame sequences: $\mathbf v^{face}=\{\mathbf{au}_{1}, \mathbf{au}_{2}, ..., \mathbf{au}_{n}\}$ and $\mathbf v^{back}=\{\mathbf b_1, \mathbf b_2, ..., \mathbf b_n\}$, where each $\mathbf{au}_{j}$ and $\mathbf b_j$ represent a learned descriptor describing the $j^{th}$ element in $\mathbf v^{face}$ and $\mathbf v^{back}$ respectively.
 %We pass both these sequences through a ResNet101 \cite{resnet} model trained on ImageNet \cite{imagenet} and concatenate both them to obtain our final visual representation $f^V=\{\mathbf{au}_{1}\oplus \mathbf{b_1}, \mathbf{au}_{2}\oplus \mathbf{b_2}, ..., \mathbf{au}_{n}\oplus \mathbf{b_n}\}$.
Two identical Bi-LSTM-based sequence representation modules, which take $\mathbf v^{face}$ or $\mathbf v^{back}$ as inputs, are employed to obtain the initial regional descriptors $\mathbf v^{face}\in \mathbb R^{n\times d^{hid}_v}$ or $\mathbf v^{back}\in \mathbb R^{n\times d^{hid}_v}$, where $d^{hid}_{v}$ is the number of the final embedding layer units in the Bi-LSTM model.  Similar to the approach followed in the text feature representation process, a stacked self-attention layer network ($F^{SA}_v$), which retrieves the multi-view attention between $\mathbf v^{face}$ and $\mathbf v^{back}$ to derive a self-attended visual descriptor  $\mathbf f^{v}\in \mathbb R^{2n\times d_{v}}$ for $V$.

\subsubsection{Audio Representation}
%Patchout fast (2-D) spectrogram transformer (PASST) \cite{passt} model, which is initialized from an ImageNet vision transformer model, and further pre-trained on 10s audio from AudioSet \cite{audioset}, is used to represent the audio component $A$ of the utterance $u$. Each segment is then represented in terms of their PASST descriptor, so $f^A=\{\mathbf a_1, \mathbf a_2, ..., \mathbf a_e\}$, where  $\mathbf a_i\in \mathbb R^{d_{A}}$ is the PASST feature of the $i^{th}$ audio frame.

Patchout fast (2-D) spectrogram transformer (PASST) \cite{passt} model, which is initialized from an ImageNet vision transformer
model, and further pre-trained on 10s audio from AudioSet \cite{audioset}, is used to represent the audio component $A$ of the utterance $u$.  
%In our experiments, we use \textit{base2level} that concatenates a longer window (160 ms and 800 ms) for timestamp embeddings. In fact, to capture the fine-grained temporal patterns within $u$, a sliding window of 100 ms and 30Hz frame rate is used to segment the audio signal. 
Each segment is then represented in terms of their PASST descriptor, so $A=\{\mathbf a_1, \mathbf a_2, ..., \mathbf a_e\}$, where  $\mathbf a_i\in \mathbb R^{d_{passt}}$ is the PASST feature of the $i^{th}$ audio frame.  Similar to visual and text representations discussed above, a Bi-LSTM network followed by an M-layered self-attention module  $F^{SA}_a$ is leveraged to obtain attention-enhanced descriptor $\mathbf f^{a}\in \mathbb R^{e\times d_{a}}$.

\subsection{Multimodal Attention Network (MAN) for Cross Attended Feature Representation}
\label{HAN_par}

While the intra-modal feature discriminability can be addressed by the proposed technique above, toward integrating the contents across modalities, heterogeneity of the data patterns across multiple modes is often a bottleneck. For example, an utterance by a speaker may reflect an impression on the speaker's face as well as on that of other participants in the conversation. Similarly, the transcript of the utterance should also be relevant to the visual background context or may have been a response to another utterance by a previous speaker. With this intuition, we propose a Multimodal cross-attended feature representation learning using a Multimodal Attention Network (\textit{MAN}) that takes $\mathbf f^{m}$ ($m\in \{v, t, a\}$) as inputs, and models cross-modal interactions at various levels of detail. As observed in Figure \ref{fig:arch}, each layer of our \textit{MAN} models cross-modal interaction by ``injecting" features from multiple \textit{Peripheral} networks into a \textit{Central} network. 

More specifically, for each mode $m$ in consideration, its mode-specific \textit{Central} query network is designed using $h$ dense layers followed by a Softmax layer\cite{krizhevsky2017imagenet}, which takes $\mathbf f^{m}$ as input and its intermediate $l^{th}$ dense layer output $\mathbf g^{l}_{m}$,  is cross-attended by one or more pairs of \textit{Peripheral} key ($\mathbf K^{l}_{m_i}$) and value ($\mathbf V^{l}_{m_i}$), such that $m_i\neq m$. Each key-value pair is generated via linear mappings. Thus, we have $\mathbf K^{l}_{m_i}=\mathbf (W^{l, K}_{m_i})^{T}\mathbf {f}^{m_i}$ and $\mathbf V^{l}_{m_i}=\mathbf (W^{l, V}_{m_i})^{T}\mathbf {f}^{m_i}$ for $\mathbf W^{l,V}_{m_i},\mathbf W^{l,K}_{m_i}\in\mathbb R^{d_{m_i}\times d_l}$. 
%The output for a single attention head $i$ for the cross-attended \textit{Central} network, having $h$ heads at its $l^{th}$ layer is computed as:
 In a multi-head attention learning framework, a particular head for the cross-attended \textit{Central} network output from its $l^{th}$ layer is computed as:

%$$K^{l, i}_{m} = \mathrm{softmax}(\mathrm{linear}(\sum_{\forall m_i \in \mathcal{M}\setminus m} K^{l}_{m_i})))$$
%$$V^{l, i}_{m} = \mathrm{softmax}(\mathrm{linear}(\sum_{\forall m_i \in \mathcal{M}\setminus m}V^{l}_{m_i})))$$
%$$\mathbf g^{l, i}_{m}=\mathbf g^{l-1}_{m}+\frac{1}{|\mathcal M|}\mathrm{softmax}(\frac{\mathbf g^{l-1}_{m}(\mathbf K^{l})^T}{\sqrt{d_{m_i}}})\mathbf V^{l}$$

\begin{equation}
\small
  \mathbf g^l_{m}=\mathbf g^{l-1}_{m}+\frac{1}{|\mathcal M|}\sum_{m_{i}\in \mathcal M\setminus \{m\} }\mathrm{softmax}(linear(\frac{\mathbf g^{l-1}_{m}(\mathbf K^{l-1}_{m_i})^T}{\sqrt{d_{m_i}}}))\mathbf V^{l-1}_{m_i}
  \end{equation}

Where $\mathcal M$ represents the set of all modes in consideration. Such responses from multiple heads are average pooled to derive the final output of the \textit{Central} network.

%\textcolor{red}{how the heads are combined, concatenated or attended? big O operator significance}

While  \textit{MAN} architecture is generic and can be extended for any number of modalities, as described in Section \ref{Uni_mode}, in our experiments, we use information from three different modes (i.e. $\mathcal M \subseteq\{t,v, a\}$). As shown in Figure 1, each mode-specific \textit{Central} query network for each $m\in \{a,v,t\}$) of  \textit{MAN}s thus produces an average pooled cross-attended mode-specific descriptor $\mathbf f^{m}_{CA}\in \mathbb R^d$ for the uni-mode components $T$, $V$, and $A$ for $u$.  We will discuss the learning algorithm later in Section \ref{learn_par}.

%Specifically, for a \textit{Central} mode $m$, at attention head $i$ of the $l^{th}$ \textit{MAN} layer, the output $g^l$ is attended by generating keys $K^l_n \forall n \in \mathcal{M}\setminus m$ and values $V^l_n \forall n \in \mathcal{M}\setminus m$ from the \textit{Peripheral} modalities through learnable linear mappings $K^{l, i}_n = W_{n, k}^{l, i}f^n$, $V^{l, i}_n = W_{n, v}^{l, i}f^n$, where $W_{n, k}^{l, k}$ and $W_{n, v}^{l, i}$ are learnable weights for each combination of modality, layer, and attention head, and $\mathcal{M}$ is the set of all modalities. The final key-value pair for an attention head $i$ is computed as a combination of all the individual key-value pairs from each \textit{Peripheral} modality as $$K^{l, i} = \sigma(d^k_i(\prod_{\forall n \in \mathcal{M}\setminus m} K^{l, i}_n)))$$ $$V^{l, i} = \sigma(d^v_i(\prod_{\forall n \in \mathcal{M}\setminus m}V^{l, i}_n)))$$ where $d^k$ and $d^v$ are dense layers, $\prod$ is the element-wise hadamard product and $\sigma$ is the softmax function.

%The final result of an \textit{MAN} layer $g^l$ for \textit{Central} modality $m$ is: $$g^l_m = B\left(g^{l-1}_m + \frac{1}{h|\mathcal{M}|}\bigoplus_{i=1}^h\mathlarger{\sigma}(\frac{g^{l-1}_mW^{l, i}_{m, q}{K^{l, i}}^T}{\sqrt{d_m}})V^{l, i}\right)$$ where $h$ is the number of attention heads and $B$ is batch normalization. As seen in Figure \ref{fig:arch}, we use $\mathcal{M}\subseteq \{t, v, a\}$ and three parallel instances of \textit{MAN} modules each with a separate \textit{Central} mode from $\mathcal{M}$ with the others as \textit{Peripheral} modes.

\begin{table*}[ht]    
\caption {\small{Performance Comparison of different methods using the weighted average F1 measure (W-Avg F1) on the MELD dataset with uni (T:=Text, A:=Audio, and V:= Video) and multi-modal Data Representations. Due to the imbalanced class distribution of the dataset, the `Fear' and `disgust' classes are represented as the minority classes, the proposed method was also compared against other $5$ majority classes (`Neutral', `Surprise', `Sadness', `Joy', and `Anger' ) in the dataset and the results are reported in column `w-avg F1 5 CLS'. `Feature Concat' in row-12 and row-13 describes the concatenation of multiple uni-mode descriptors to define a multimodal descriptor.}}
\label{tab:meld}
\smaller{
      \centering
    \begin{tabular}{|c|c|c|c|c|c|c|c|c|c|c|}
    \hline
        \multirow{2}*{Method} & \multirow{2}*{Mode} & \multirow{2}*{Neutral} & \multirow{2}*{Surprise} & \multirow{2}*{Fear} & \multirow{2}*{Sadness} & \multirow{2}*{Joy} & \multirow{2}*{Disgust} & \multirow{2}*{Anger} & \multirow{2}*{w-Avg F1} & w-avg F1 \\
         & & & & & & & & & & 5-CLS \\ \hline
       \makecell{MFN\cite{zadeh2018memory}} & T + A & 0.762 & 0.407 & 0.0 & 0.137 & 0.467 & 0.0 & 0.408 & 0.547 & 0.5732\\ \hline
       \multirow{3}*{\makecell{ICON\cite{hazarika-etal-2018-icon}}} & T & 0.762 & 0.462 & 0.0 & 0.189 & 0.485 & 0.0 & 0.301 & 0.546 & 0.5718\\ \cline{2-11}
        & A & 0.669 & 0.0 & 0.0 & 0.0 & 0.086 & 0.0 & 0.315 & 0.377 & 0.3947\\ \cline{2-11}
        &  T + A & 0.736 & 0.500 & 0.0 & 0.232 & 0.502 & 0.0 & 0.448 & 0.563 & 0.5897\\ \hline
         \multirow{3}*{\makecell{DialogueRNN \cite{majumder2019dialoguernn}} } & T & 0.737 & 0.449 & 0.054 & 0.234 & 0.476 & 0.0 & 0.415 & 0.551 & 0.5759\\ \cline{2-11}
         & A & 0.53 & 0.156 & 0.0 & 0.083 & 0.112 & 0.051 & 0.321 & 0.34 & 0.3542\\ \cline{2-11}
        &  T + A & 0.732 & 0.519 & 0.0  & 0.248 &0.532 & 0.0 & 0.456  & 0.57 & 0.5971\\ \hline
        \multirow{3}*{\makecell{ConGCN  \cite{ijcai2019-752}} }& T & 0.749 & 0.498 & 0.065 & 0.226 & 0.524 & 0.088 & 0.432 & 0.574 & 0.5969\\ \cline{2-11}
        & A & 0.641 & 0.254 & 0.047 & 0.193 & 0.155 & 0.030 & 0.341 & 0.422 & 0.44\\ \cline{2-11}
        &  T + A & 0.767 & 0.503 & 0.087 & 0.285 & 0.531 & 0.106 & 0.468 & 0.594 & 0.6175\\ \hline
        \makecell{DialogueCRN \cite{hu2021dialoguecrn}}& T + A & - & - &- &- &- &- &- & 0.6073 &-\\ \hline
       % \makecell{ERLDK\\ \cite{9400391}}& T + A + V & - & - &- &- &- &- &- & 0.5972 &-\\ \hline
        \makecell{EmoCaps \cite{li2022emocaps}} & T + A + V & 0.7712 & 0.6319 &0.0303 &0.4254 &0.5750 &0.0769 &0.5754 & 0.6400 &-\\ \hline
        \makecell{M2FNet \cite{chudasama2022m2fnet}} & T + A + V & - & - &- &- &- &- &- & 0.6785 &-\\ \hline
        \makecell{Cross-Modal Distribution Matching \cite{liang2020semi} }& T + A & - & - &- &- &- &- &- & 0.571 &-\\ \hline
        \makecell{Transformer Based  Cross-modality Fusion \cite{xie2021robust} }& T + A +V & - & - &- &- &- &- &- & 0.64 &-\\ \hline        
        \makecell{Hierarchical Uncertainty \\for Multimodal Emotion Recognition \cite{chen2022modeling} }& T + A +V & - & - &- &- &- &- &- & 0.59 &-\\ \hline   
        \makecell{Shape of Emotion \cite{agarwal2021shapes} }& T + A +V & - & - &- &- &- &- &- & 0.63 &-\\ \hline 
        \makecell{UniMSE\cite{hu2022unimse} }& T + A +V & - & - &- &- &- &- &- & 0.66 &-\\ \hline
        %\makecell{Multimodal Attentive Learning\\ \cite{arumugam2022multimodal}}& T + A + V & 0.821 & 0.6276 & \textbf{0.1429} & 0.3947 & 0.6402 & \textbf{0.0988} & 0.5997 & 0.6773 & 0.7038 \\ \hline
      \multirow{3}*{\makecell{Proposed Uni-mode \\ Feature Rep. (Section \ref{Uni_mode})\\+Classifier (Section \ref{spa_temp})}} 
        & T & 0.7439 & 0.6191 & 0.0209 & 0.3914 & 0.5178 & 0.0613 & 0.5036 & 0.6041 & 0.6306\\ \cline{2-11}
        & A & 0.3838 & 0.3581 & 0.0209 & 0.3286 & 0.3617 & 0.0613 & 0.3529 & 0.3537 & 0.3684\\ \cline{2-11}
        & V & 0.5562 & 0.4905 & 0.0209 & 0.3374 & 0.4098 & 0.0613 & 0.3713 & 0.4615 & 0.4813\\ \cline{2-11}\hline
        \multirow{4}*{\makecell{Proposed Uni-mode \\ Feature Rep.(Section \ref{Uni_mode}) + Feature Concat.\\+ Classifier (Section \ref{spa_temp})}} & T + A & 0.7627 & 0.6318 & 0.0241 & 0.4214 & 0.5316 & 0.0613 & 0.5597 & 0.6265 & 0.6540 \\ \cline{2-11}
        & T + V & 0.7427 & 0.6218 & 0.0241 & 0.4214 & 0.5316 & 0.0613 & 0.5597 & 0.6158 & 0.6428\\ \cline{2-11}
        & A + V & 0.5562 & 0.5796 & 0.0209 & 0.3610 & 0.4098 & 0.0613 & 0.4318 & 0.4810 & 0.5017\\ \cline{2-11}
        & T + A + V & 0.7671 & 0.6518 & 0.0319 & 0.4629 & 0.5291 & 0.0691 & 0.5713 & 0.6356 & 0.6632 \\ \cline{2-11}\hline
        \makecell{Proposed \textit{MAN}-based \\ Feature Rep.(Section \ref{HAN_par}) + Feature Concat\\+Classifier (Section \ref{spa_temp})\\} &  T + A + V & 0.8359 & 0.7094 & 0.0674 & 0.4468 & 0.6297 & 0.0891 & 0.6389 & 0.6992 & 0.7286\\ \cline{2-11}\hline
        \textit{AMuSE} & T + A + V  & \textbf{0.8469} & \textbf{0.7283} & 0.0674 & \textbf{0.4632} & \textbf{0.6481} & 0.0891 & \textbf{0.6574} & \textbf{0.7132} & \textbf{0.7431}\\ \hline
    \end{tabular}}
    \vspace{-5mm}
\end{table*}

%\subsection{Temporal Feature Processing}
%Once we obtain the cross-attended ($\mathbf c^{m}, m\in \mathcal M$) descriptors obtained from the respective \textit{MAN} modules, we process the temporal features of these descriptors via a Bi-directional-Long Short-Term Memory (Bi-LSTM) to obtain the mode-specific representation ($\mathbf r^{m} \in \mathbb{R}^{e} \forall m \in \mathcal M$) for $u$.
%In contrast, we make two observations in this context: (1) The importance of all modes is not always uniform or constant. In fact, they do vary across samples from different categories. In other words, while human affective states are expressed via both verbal and non-verbal channels in parallel, their mutual interactions, and relative dynamics are still not consistent across several emotion categories. A lack of understanding of this mutual interplay may generate the risk of misclassification; (2) The transition of an individual's expression of emotion is continuous. Therefore, slight changes in the utterance representations should not have caused drastic changes in the utterance's emotion labels.

\vspace{-2mm}
\subsection{\textit{A}daptive \textit{F}usion (AF)}
%TODO: Reword observations and explain how they're relevant to the fusion we propose
Most of the existing methods \cite{majumder2019dialoguernn,hu2021dialoguecrn,arumugam2022multimodal} leverage a static approach for multimodal feature fusion. Here we make an important observation: The relative importance of each modality is not uniform and varies across samples exhibiting different emotions. For example, a speaker's emotion may reflect an influence of several contexts like audience reaction or surrounding environment. However, the transition of an individual's expression of emotion is continuous. Therefore, slight changes in the utterance representations should not have caused drastic changes in the utterance's emotion labels. Keeping this in mind, we propose an \textit{A}daptive \textit{F}usion (\textit{AF}) function $\mathcal{A}$ that learns a linear combination of the mode-specific representations ($\mathbf f^{m}_{CA}$), to derive a comprehensive \textit{spatial multimodal descriptor} $\mathcal{A}(u)$ for an utterance $u$ as follows:
\begin{equation}
\label{eq:amus}
    \mathcal A(u) = \bigoplus_{m \in \mathcal{M}}\left(\frac{1}{|\mathcal{M}|}\sum_{m_i\in\mathcal M\setminus m}\alpha^m_{m_i}\mathbf f^{m_i}_{CA}+ (1 - \alpha^m_{m_i})\mathbf f^{m_i}_{CA}\right)
\end{equation}
where $0 \leq\alpha^m_{m_i} \forall m,m_i \in \mathcal M\leq 1$ are learnable parameters and $\bigoplus$ represents the concatenation operator. Thus, the proposed fusion function $\mathcal A$ provides a flexible multimodal representation mechanism, by which the resulting multimodal descriptor $\mathcal A(u)$ for an utterance $u$ is able to retain category-specific discriminative data patterns, however not completely disregarding the unique instance-specific data patterns observed in the utterance.

\subsection{Emotion Classification}
\label{spa_temp}

Given a conversational dialogue represented using a sequence of $n$ utterances $\{u_j\}^{n}_j\in \mathcal D$, our task is to evaluate the emotion of a user-identified speaker $s$ by utilizing the spatio-temporal contexts observed in the dialogue. To attain this objective,  we design a speaker-specific representation of the dialogue by using two parallel utterance sequences: \textit{Dialogue Context}, which describes the entire sequence $\{\mathcal A(u_j)\}_j$; \textit{Speaker Context}, a derived utterance sub-sequence $\{\mathcal A(u_{{s}_j})\}_j$, where the sub-sequence $\{u_{{s}_j}\}_{{s_{j}\in [1, n]}}$ is generated from the dialogue and includes only those utterances, in which $s$ vocally contributes to the conversation. Given the voice of a speaker identified by the user in the first keyframe, we use the Librosa library function to match it across utterances to identify this derived subsequence. Two parallel Bi-LSTMs are trained to capture the spatio-temporal contexts independently from these contexts' perspectives.  In our experiments, Bi-LSTMs were found to be more useful than LSTm due to their ability to capture the bidirectional temporal contexts.
For example using the \textit{Speaker Context}, the pertinent spatio-temporal representation $\mathbf s_l \in \mathbb R^{s}$ of an utterance $u^{s}_l$ is reasoned from the representation $\mathbf s_{(l-1)}$ of $u^{s}_{(l-1)}$, while also considering the current state of utterance $u^{s}_{l}$ denoted as $\mathbf z^{s}_{l}$ (which are initially null) - as $\mathbf s_{l}= \overleftrightarrow{\mathrm{LSTM}^{s}}(\mathbf s_{(l-1)},\mathbf z^{s}_{l})$. Similarly,  using the \textit{Dialogue Context}, the pertinent spatio-temporal representation $\mathbf d_l \in \mathbb R^{s}$ of an utterance $u_l$ is reasoned from the representation $\mathbf d_{(l-1)}$ of $u_{(l-1)}$, while also considering the current state of utterance $u_{l}$ denoted as $\mathbf w^{d}_{l}$ (which are initially null) - as $d_{l}= \overleftrightarrow{\mathrm{LSTM}^{d}}(\mathbf d_{(l-1)},\mathbf w^{s}_{l})$.
The final representation of each utterance in the \textit{Speaker Context} is then derived as, $e_l = \mathbf s_{l} \oplus \mathbf d_{l}$, which is used as an input to a simple neural network comprising of a linear layer followed a softmax activation to estimate the occurrence of emotion in the utterance.

\subsection{Learning}
\label{learn_par}
The learning algorithm of \textit{AMuSE} includes two independent learning objectives. A loss objective ($\mathcal{L})$ for \textit{MAN} learning and an \textit{AF} objective to optimize the values of $\alpha^m_{m_i}$.

\subsubsection{MAN Learning}
Given an utterance $u_{j}\in \mathcal D$, the proposed Multimodal Attention Network (\textit{MAN}) jointly learns the cross-attended representations $\mathbf f^{t}_{CA,j}$, $\mathbf f^{v}_{CA,j}$, and $\mathbf f^{a}_{CA,j}$ with twofold objectives: 1) preserving Instance-level discriminability \cite{gutmann2010noise,oord2018representation} within them. This is equivalent to obtaining the learned representations such that they must be more discriminative of $u$ compared to the other samples in $\mathcal D$. To incorporate this intuition, we use Noise Contrastive
Estimation (NCE) \cite{gutmann2010noise} loss ($\mathcal L_{NCE}$); 2) preserving the `category-level' information within their learned representations, such that the predictions obtained from each of these learned mode-specific representations may also align with the ground-truth labels. We leverage Focal Loss \cite{focal} for this purpose.

We leverage an aggregated noise contrastive estimation ($\mathcal L_{ACE}$) and an averaged focal loss ($\mathcal L_{fl}$), as defined below:
\begin{equation}
\label{fl_loss}
    \mathcal{L_{ACE}}=\frac{1}{|\mathcal D|}\sum_{u_{j}\in \mathcal D}\frac{1}{|\mathcal M|}\sum_{\substack {m\neq m_{i}\\ m, m_{i}\in \mathcal M }}{\mathcal{L}_{\mathrm{NCE}}(\mathbf f^{m}_{CA,j}, \mathbf f^{m_i}_{CA,j})} 
\end{equation}
with

% \begin{multline}
% \mathcal{L}_{\mathrm{NCE}}(\mathbf f^{m}_{CA,j}, \mathbf f^{m_{i}}_{CA,j})=\big[-log(\frac{P(\mathbf f^{m_i}_{CA,j}|\mathbf f^{m_i}_{CA,j})}{P(\mathbf f^{m_i}_{CA,j}|\mathbf f^{m_i}_{CA,j})+\frac{|\mathcal N_j|}{|\mathcal N|}}) + \\
% \nbsp \sum_{k\in \mathcal N_{j}}log(\frac{P(\mathbf f^{m_i}_{CA,k}|\mathbf f^{m_i}_{CA,j})}{P(\mathbf f^{m_i}_{CA,k}|\mathbf f^{m_i}_{CA,j})+\frac{|\mathcal N_j|}{|\mathcal N|}})-1\big]
% \end{multline}

\begin{multline}
\mathcal{L}_{\mathrm{NCE}}(\mathbf{f}^{m}_{CA,j}, \mathbf{f}^{m_{i}}_{CA,j}) = \\
\Bigg[-\log\Bigg(\frac{P(\mathbf{f}^{m_i}_{CA,j}|\mathbf{f}^{m_i}_{CA,j})}{P(\mathbf{f}^{m_i}_{CA,j}|\mathbf{f}^{m_i}_{CA,j}) + \frac{|\mathcal{N}_j|}{|\mathcal{N}|}}\Bigg) + \\
\sum_{k\in \mathcal{N}_{j}}\log\Bigg(\frac{P(\mathbf{f}^{m_i}_{CA,k}|\mathbf{f}^{m_i}_{CA,j})}{P(\mathbf{f}^{m_i}_{CA,k}|\mathbf{f}^{m_i}_{CA,j}) + \frac{|\mathcal{N}_j|}{|\mathcal{N}|}}\Bigg) - 1\Bigg]
\end{multline}

that computes the 
probability of both features $\mathbf f^{m}_{CA,j}$ and $\mathbf f^{m_i}_{CA,j}$ representing the same instance $u_j$ compared to other elements in a uniformly sampled negative set $\mathcal N_j$.

The averaged Focal Loss \cite{focal}, specifically effective for an imbalanced dataset like ours, is defined below:
\begin{equation}
    \mathcal{L}_{fl} = \frac{1}{|\mathcal M|}\frac{1}{|\mathcal D|} \sum_{m\in \mathcal M}\sum_{u_j\in \mathcal D}(1-p^{m}_{c}(u_{j}))^\gamma p^{m}_{c}(u_{j})
\end{equation}
where $p^{m}_{c}(u_{j})$ is the predicted class-membership probability score for the sample $u_j$ by the $m^{th}$ mode-specific \textit{Central} query network and $\gamma$ is a tunable parameter. We use a combined loss function $\mathcal L=\mathcal L_{ACE}+\mathcal L_{fl}$ to jointly learn its mode-specific \textit{Central} query networks.

\begin{table*}[ht]
\caption {Performance comparison of difference methods using the weighted average F1 measure (W-Avg F1) on the IEMOCAP dataset with with uni (T:=Text, A:=Audio, and V:= Video) and multi-modal Data Representations. `Feature Concat' in row 13 and row 14 describe the concatenation of multiple uni-mode descriptors to define a multimodal descriptor.}
\label{tab:iemocap}
\centering
\begin{tabular}{|c|c|c|c|c|c|c|c|c|}
    \hline
       \textbf{Method} & \textbf{Mode} &  \textbf{Happy} &  \textbf{Sad} &  \textbf{Neutral} &  \textbf{Angry} &  \textbf{Excited}  &  \textbf{Frustrated} &  \textbf{w-Avg F1}\\     \hline
      \makecell{MFN\cite{zadeh2018memory}} & T + A  & - & - & - & - & - & - & 0.3490\\ \hline
      \makecell{ICON\cite{hazarika-etal-2018-icon}} & T + A + V & 0.3280 & 0.7440 & 0.6060 & 0.6820 & 0.6840 & 0.6620 & 0.6350\\ \hline 
      \makecell{DialogueRNN\cite{majumder2019dialoguernn}} & T + A + V & 0.3318 & 0.7880 & 0.5921 & 0.5128 & 0.7186 & 0.5891 & 0.6275\\ \hline
      \makecell{MMGCN\cite{10.1145/3343031.3351034}} & T + A + V & 0.4235 & 0.7867 & 0.6173 & 0.6900 & 0.7433 & 0.6232 & 0.6622\\ \hline
      \makecell{DialogueCRN\cite{hu2021dialoguecrn}} & T + A  & 0.6261 & 0.8186 & 0.6005 & 0.5849 & 0.7517 & 0.6008 & 0.6620\\ \hline
      \makecell{ERLDK\cite{9400391}} & T + A & 0.4730 & 0.7919 & 0.5642 & 0.6054 & 0.7444 & 0.6385 & 0.6390\\ \hline
              \makecell{Hierarchical Uncertainty \\ for Multimodal Emotion Recognition \cite{chen2022modeling} }& T + A + V & -& - & -& - & -& - & 0.6598\\ \hline
%      \makecell{EmoCaps\\\cite{9400391}} & T + A + V & \textbf{0.7191} & \textbf{0.8506} & 0.6448 & \textbf{0.6899} & 0.7841 & 0.6676 & 0.7177\\ \hline
      \makecell{DAG-ERC+HCL\cite{Yang2021HybridCL}} & T  & - & - & - & - & - & - & 0.6803\\ \hline
       \makecell{M2FNet\cite{chudasama2022m2fnet}} & T + A + V & - & - & - & - & - & - & 0.6986\\ \hline
       \makecell{Multimodal Attentive Learning\cite{arumugam2022multimodal}} & T + A + V & - & - & - & - & - & - & 0.6540\\ \hline
        \multirow{3}*{\makecell{Proposed Uni-mode \\ Feature Rep. \\(Section \ref{Uni_mode})\\}} 
            & T  & 0.2991 & 0.6141 & 0.5251 & 0.5728 & 0.5918 & 0.5969 & 0.5526\\ \cline{2-9}
            & A & 0.2991 & 0.3894 & 0.3951 & 0.2749 & 0.326 & 0.3316 & 0.3417\\ \cline{2-9}
            & V & 0.3038 & 0.5329 & 0.5619 & 0.2749 & 0.326 & 0.431 & 0.4260\\ \cline{2-9}\hline
           \multirow{4}*{\makecell{Proposed Uni-mode \\ Feature Rep.\\(Section \ref{Uni_mode})\\+ Feature Concat.\\}} & T + A & 0.3038 & 0.6368 & 0.5619 & 0.598 & 0.6027 & 0.6069 & 0.5727\\ \cline{2-9}
            & T + V & 0.3359 & 0.6368 & 0.5885 & 0.598 & 0.6027 & 0.6069 & 0.5815\\ \cline{2-9}
            & A + V & 0.3038 & 0.5592 & 0.6328 & 0.321 & 0.326 & 0.5293 & 0.4782\\ \cline{2-9}
            & T + A + V & 0.3917 & 0.6368 & 0.6354 & 0.6374 & 0.6027 & 0.6399 & 0.6117\\ \cline{2-9}\hline
           \makecell{Proposed  \textit{MAN}-based \\ Feature Rep.(Section \ref{HAN_par})+ Feature Concat\\}  & T + A + V & 0.6591 & 0.8106 & 0.7248 & 0.6599 & 0.7769 & 0.6734 & 0.7147\\ \cline{2-9}\hline
            \textit{AMuSE} & T + A + V & \textbf{0.7025} & \textbf{0.8418} & \textbf{0.7548} &\textbf{ 0.6748} & \textbf{0.7935} & \textbf{0.6923} & \textbf{0.7391}\\ \hline
    \end{tabular}
    \vspace{-5mm}
    \end{table*} 

\vspace{-1.2mm}
\subsubsection{AF Learning}
\label{learn_amuf}
% As observed in Eqn.\ref{eq:amus}, the multimodal descriptor $\mathcal A(u)$ interpolates all cross-attended mode-specific descriptors ($\mathbf{F}^{m_i}_{CA}$) to reveal all the discriminative feature information by leveraging the changes in the model behavior in response to varying inputs. As such, it is intuitive to note that a slight change in the feature representation should not cause any observable change in the model's decision.  Nevertheless, manual selection for any of the interpolation coefficients (however small it is) $\boldsymbol{\alpha}^{u}_{m_i}$ may not be equally effective across all samples. Thus, the threefold approximation task specific to our scenario is solved in a pairwise manner. In other words, we perform the learning of these interpolation parameters by first approximating $\boldsymbol{\alpha}_{1}^'$ for $\mathbf u^{'}_{1}=\boldsymbol{\alpha}_{1}^{'}\mathbf{f}^{m_1}_{CA}+(1-\boldsymbol{\alpha}_{1}^{'})\mathbf f^{m_2}_{CA}$ followed by approximating $\boldsymbol{\alpha}_{2}^'$ for $\mathbf u^{'}_{2}=\boldsymbol{\alpha}_{2}^{'}\mathbf u^{'}_{1}+(1-\boldsymbol{\alpha}_{2}^{'})\mathbf f^{m_3}_{CA}$. Then by setting $F^{AMuS}(u)=\frac{1}{3}(\mathbf u^{'}_{1}+\mathbf u^{'}_{2})$ and equating the coefficients of like terms, we obtain 
As observed in Eqn.~\ref{eq:amus}, the multimodal descriptor $\mathcal{A}(u)$ interpolates all cross-attended mode-specific descriptors ($\mathbf{F}^{m_i}_{CA}$) to reveal all the discriminative feature information by leveraging the changes in the model behavior in response to varying inputs. As such, it is intuitive to note that a slight change in the feature representation should not cause any observable change in the model's decision. Nevertheless, manual selection for any of the interpolation coefficients (however small it is) $\boldsymbol{\alpha}^{u}_{m_i}$ may not be equally effective across all samples. Thus, the threefold approximation task specific to our scenario is solved in a pairwise manner. In other words, we perform the learning of these interpolation parameters by first approximating $\boldsymbol{\alpha}_{1}^{'}$ for $\mathbf{u}^{'}_{1}=\boldsymbol{\alpha}_{1}^{'}\mathbf{f}^{m_1}_{CA}+(1-\boldsymbol{\alpha}_{1}^{'})\mathbf{f}^{m_2}_{CA}$ followed by approximating $\boldsymbol{\alpha}_{2}^{'}$ for $\mathbf{u}^{'}_{2}=\boldsymbol{\alpha}_{2}^{'}\mathbf{u}^{'}_{1}+(1-\boldsymbol{\alpha}_{2}^{'})\mathbf{f}^{m_3}_{CA}$. Then by setting $F^{AMuS}(u)=\frac{1}{3}(\mathbf{u}^{'}_{1}+\mathbf{u}^{'}_{2})$ and equating the coefficients of like terms, we obtain $\boldsymbol{\alpha}^{u}_{m_1}=\boldsymbol{\alpha}^{'}_{1}\cdot\boldsymbol{\alpha}^{'}_{2}$, $\boldsymbol{\alpha}^{u}_{m_2}=\boldsymbol{\alpha}^{'}_{2}\cdot(1-\boldsymbol{\alpha}^{'}_{1})$, and $\boldsymbol{\alpha}^{u}_{m_3}=1-\boldsymbol{\alpha}^{'}_{2}$. For optimizing the interpolation parameter $\boldsymbol{\alpha}_{1}^{'}$ (and similarly $\boldsymbol{\alpha}_{2}^{'}$), we adopt the optimization approach of \cite{alfamix}, which is as follows:
\begin{equation} \label{eq:alfamix}
   \boldsymbol{\alpha}_{1}^{'}\approx\epsilon\frac{\lVert(\mathbf f^{m_1}_{CA} - \mathbf f^{m_2}_{CA})\rVert_2\nabla_{\mathbf f^{m_2}_{CA}}\mathcal{L}(Q_{m_{2}}(\mathbf f^{m_2}), y^*)}{\lVert{\nabla{\mathbf f^{m_2}_{CA}}\mathcal{L}(Q_{m_{2}}(\mathbf f^{m_2}), y^*)}\rVert_2}\oslash (\mathbf f^{m_1}_{CA}-\mathbf f^{m_2}_{CA})
\end{equation}
where $y^*$ is the ground truth label for the sample $u$, $\oslash$ is the element-wise division, $\epsilon$ is the hyper-parameter that controls the amount of interpolation in the result, and $Q_{m_{i}}$ represents the $i^{th}$ mode-specific \textit{Central} query network. To facilitate the learning process, we randomly identify a set of informative samples from the validation pool for which the loss due to a small interpolation is indeed affected (i.e. system prediction indeed changes by a slight change in the interpolation parameters) to use in the following training epochs.

\vspace{-2mm}
\section{Experiments}
\subsection{Datasets}
%TODO: Expand on the datasets
Derived from the TV series F.R.I.E.N.D.S, \textit{MELD} \cite{poria2019meld} is a multi-party multimodal conversation dataset comprising 7 emotions - `Anger', `Disgust', `Sadness', `Joy', `Surprise', `Fear', and `Neutral'. \textit{IEMOCAP} \cite{Busso2008} is a dyadic conversational dataset, with recordings of professional actors performing scripted and improvised scenarios comprising 6 emotions - 'Happy', 'Sad', 'Neutral', 'Angry', 'Excited', 'Frustrated'. 

\subsection{Results \& Comparative Study}
\begin{figure*}[t]
\centerline{\includegraphics[width=0.95\textwidth]{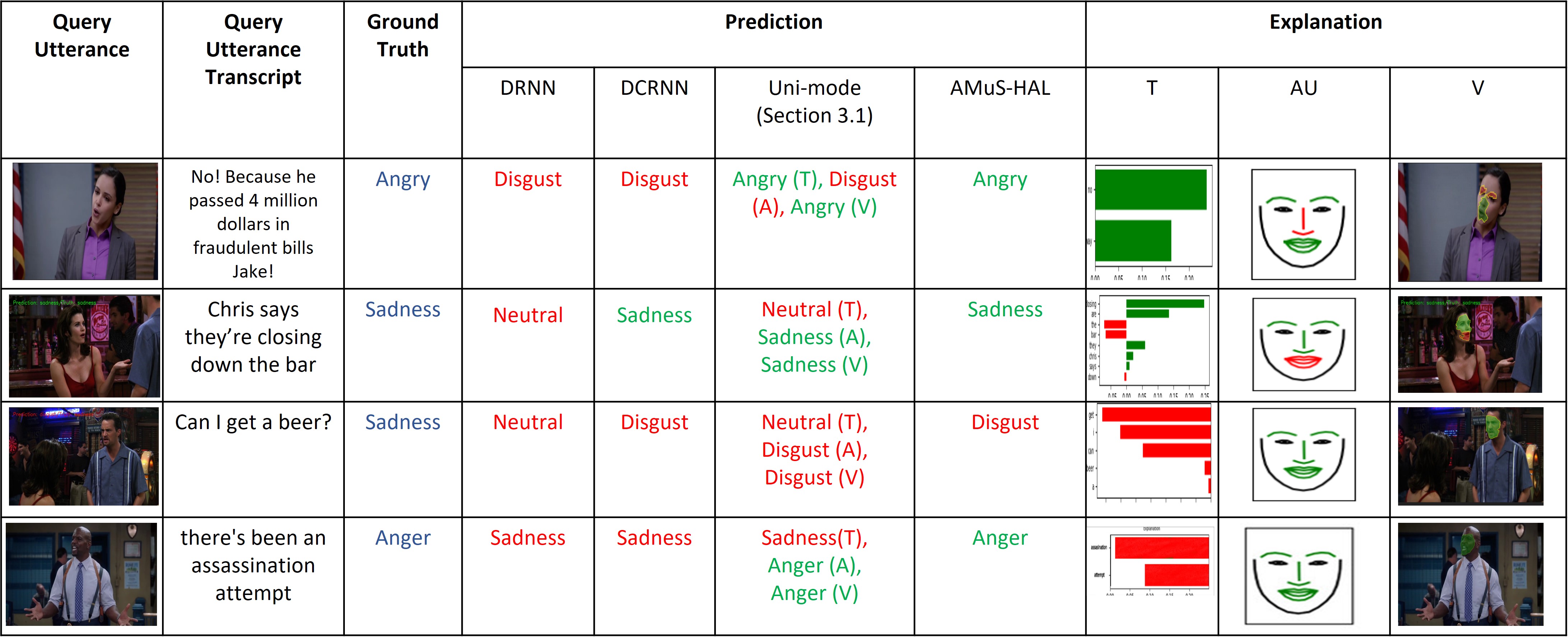}}
\caption{\small{Some example results with 3 mode-specific explainability analysis, wherein explanation columns regions/texts contributing to the model decision are highlighted in \textcolor{green}{Green}. The detracting regions are highlighted in \textcolor{red}{Red}}}
\label{fig:explain}
\vspace{-6mm}
\end{figure*}

Figure \ref{fig:explain} provides some qualitative results. Table \ref{tab:meld} and Table \ref{tab:iemocap} present the results on MELD and IEMOCAP test sets, respectively by using F1-score \cite{majumder2019dialoguernn} as the evaluation metric. The results are compared against several state-of-the-art algorithms \cite{hazarika-etal-2018-icon,zadeh2018memory,majumder2019dialoguernn,ijcai2019-752,hu2021dialoguecrn,9400391,li2022emocaps,chudasama2022m2fnet,liang2020semi,Yang2021HybridCL,arumugam2022multimodal}. For each emotion category, we also evaluate the classification performance using their weighted averages across all emotion classes. As observed in Table \ref{tab:meld}, while `Text' appears to be the most reliable uni-modal feature, combining information from multiple modes is always helpful. To this end, as we compare the last sub-row of row-12 and row-13, the proposed  \textit{MAN} based cross-attention appears to be extremely beneficial in improving the weighted 
F1-score \textbf{(w-avg F1) by around $6\%$}. Finally, using a flexible and efficient fusion approach, the proposed \textit{AMuSE} facilitates further improvement in the performance by reporting \textbf{$\sim 74\%$ w-avg F1- yet another $\sim 2\%$ improvement} compared to the results reported in the baseline row-13 scenario. Row-13 reports the experiment results, wherein cross-attended mode-specific feature descriptors (Section \ref{Uni_mode}) are simply fused using equal values of the interpolation parameters in Eqn \ref{eq:amus} (i.e. $\boldsymbol{\alpha}^{s_{1}}_{m_i} = \boldsymbol{\alpha}^{s_{2}}_{m_j}\forall m_{i},m_{j}\in \mathcal M,\forall u^{s_{1}},u^{s_{1}}\in \mathcal D$). As we compare this performance (in row-14) with row-8 and row-9 of Table \ref{perf_MELD}, we observe that \textbf{\textit{AMuSE} reports around $4-7\%$ improved performance} compared to the best performing existing methods \cite{li2022emocaps,chudasama2022m2fnet}. A similar performance pattern is also observed in Table \ref{tab:iemocap}, wherein \textit{AMuSE} is compared against several baseline methods using the IEMOCAP dataset. By comparing the last sub-row of row-13 and row-14, we find that the proposed  \textbf{\textit{MAN}-based cross-attention enables the mode to attain an impressive $10\%$ improvement} over its baseline test scenario, in which only the mode-specific feature descriptors (Section \ref{Uni_mode}) are simply concatenated to define a multimodal descriptor. Finally, by employing \textit{AF} for feature fusion, \textbf{the model attains $\sim 74\%$ w-avg F1}, which overshoots some of the best-performing baselines \cite{li2022emocaps,chudasama2022m2fnet} \textbf{ by around $2-4\%$}. While most of the works use F1-score as the evaluation metric, compared to a handful few recent works \cite{9400391,hu2021dialoguecrn}, which have also reported classification accuracy of their method, proposed \textit{AMuSE} reports an impressive performance. Compared to one of the best-performing baselines M2FNet \cite{chudasama2022m2fnet} that reports $66.71$ accuracy in the MELD dataset and $69.69\%$ accuracy in the IEMOCAP dataset, \textit{AMuSE} reports  \textbf{\textit{around $7\%$ (i.e. $73.28\%$ accuracy score in MELD dataset) and $5\%$ (i.e. $74.49\%$ accuracy score in IEMOCAP dataset) improvement respectively}}.

\vspace{-2mm}
\subsection{Ablation Study}
% \vspace{-6mm}
As observed in Table \ref{amus_able}, compared to the other testing scenarios (where either the $\{\boldsymbol{\alpha}^{u}_{m_i} \forall m_{i}\in \mathcal M,\forall u\in \mathcal D\}$ parameters were chosen at random or were fixed to the same value for all samples in $\mathcal D$ and modes $\mathcal M $) the proposed \textit{AMuSE} shows an improved performance in Test-3 experiment setting, wherein it leverages the learning algorithm for the \textit{AF} interpolation parameters, introduced in Section \ref{amus_par}, to optimize the choices of these parameters in mode-specific and input-specific manner. This makes the model more adaptable to the newer data patterns, observed in analyzing speakers' emotions from diverse socio-racial backgrounds, compared to those available in the training collection. In the other set of ablation study experiments, we choose different values for the tunable parameter $\gamma$ in the focal loss function defined in Eqn. \ref{fl_loss}. Again as observed in Table \ref{gamma_able}, in both datasets, the chosen value of $\gamma =1$ produces a slightly better W-Avg F1 score, compared to the other values of $\gamma$. In fact, the performance of \textit{AMuSE} remains mostly stable over a range of values in $[0.75, 1.25]$, which highlights the system stability in the performance over the choice of $\gamma$ values. Finally, in Table \ref{amus_layers} we also report ablation study results for the number of \textit{MAN} layers in the model. The performance remains fairly consistent when using 3, 4, or 5 layers and peaks at 4, which is the number of layers we have chosen in the model.

%\subsection{Implementation}
%For both datasets, we sample 32 input frames at 10fps resized to a size of 448x448. We do a random resize crop of these frames to 224x224, followed a by random horizontal flip with a probability of 0.5 and color augmentation. We sample stereo audio in sync with the video frames at 48kHz. Both audio and video are normalized between [-1, 1]. To generate audio features using PASST, we use a 100ms sliding window and 30Hz frame rate. For text, we cap the sentence size at 48 words. The feature representation sizes are $d^T = 768$, $d^A = d^V = 2048$ In the model, we use 4 \textit{MAN} layers in each mode-specific \textit{MAN} module. Each \textit{MAN} layer uses $h = 16$ attention heads. The mode-specific embedding size $e$ is $1024$. We choose $r = 512$ for the situation and speaker-level context representation. All linear layers in the \textit{MAN} module use the GeLU activation, while other linear layers use the ReLU activation. The model was trained using the Adam optimizer with half-period cosine annealing for learning rate scheduling ranging from 1e-4 to 1e-5 for 750 epochs with 5000 warmup steps.

\begin{table}
\caption {Ablation Study on the \textit{AF} function parameters ($\boldsymbol{\alpha}^{u}_{m_i}$ for $m_{i}\in \mathcal M$) was performed in several testing scenarios:\textit{Test-1} in which we choose these parameters at random such that $\boldsymbol{\alpha}^{u}_{m_i}\neq \boldsymbol{\alpha}^{u}_{m_j}\forall m_{i},m_{j}\in \mathcal M$ and $\boldsymbol{\alpha}^{u^{s_{1}}}_{m_i}=\boldsymbol{\alpha}^{u^{s_{2}}}_{m_i}\forall u^{s_{1}},u^{s_{2}}\in \mathcal D$; \textit{Test-2} in which we choose these parameters such that $\boldsymbol{\alpha}^{s_{1}}_{m_i} = \boldsymbol{\alpha}^{s_{2}}_{m_j}\forall  m_{i},m_{j}\in \mathcal M,\forall u^{s_{1}},u^{s_{1}}\in \mathcal D$; \textit{Test-3} in which we learn the parameters following the approach (\textit{AF Learning}) discussed in Section \ref{learn_amuf}. The table reports the weighted average F1 measure (W-Avg F1) over all classes in the datasets.}
\label{amus_able}
\centering
\begin{tabular}{|c|c|c|c|}
    \hline
       \textbf{Dataset} &  \textit{Test-1} &  \textit{Test-2} &  \textit{Test-3}\\\hline
       MELD  \cite{poria2019meld}& $70.86$ & $71.10$ & $71.32$\\\hline
       IEMOCAP \cite{Busso2008} & $72.07$ & $72.89$ &  $73.91$\\
       \hline
       \end{tabular}
       \vspace{-5mm}
    \end{table} 

\begin{table}
\caption {Ablation Study on the tunable parameter $\gamma$ in the focal loss function defined in Eqn. \ref{fl_loss}.The table reports the weighted average F1 measure (W-Avg F1) over all classes in the datasets.}
\label{gamma_able}
\centering
\begin{tabular}{|c|c|c|c|c|}
    \hline
       \textbf{Dataset} &  $\gamma=0.5$ & $\gamma=0.75$ &  $\gamma=1.0$ & $\gamma=1.25$ \\\hline
       \makecell{MELD   \cite{poria2019meld}}& $70.71$ & $71.08$ & $71.32$\ & $70.97$   \\\hline
       \makecell{IEMOCAP \cite{Busso2008}} &$71.46$ &$73.47$ &$73.91$ & $73.12$  \\
       \hline
    \end{tabular}
     \vspace{-5mm}
    \end{table}

\subsection{Explainability}
The proposed explainability analysis approach uses Local Interpretable Model-Agnostic Explanations (LIME)\footnote{\url{https://github.com/marcotcr/lime}} to explain system decisions. LIME provides Interpretable, Model-Agnostic Visual explanations for any classifier by treating the classification model as a black box. LIME approximates the classifier model locally in the neighborhood of the prediction.
In Figure \ref{fig:explain}, we present the explainability analysis in various modes: Textual explanation with the words that contribute the most (or against) the prediction; face landmarks and Action Unit (AU) based explanation that illustrates the regions in the speaker's face that contribute (using \textcolor{green}{Green}) and distract (using \textcolor{red}{Red}) to the prediction. Finally, we present visual regions of interest in the image responsible for the model's decision.

\begin{table}
  \centering
    \caption{The ablation study showing the effect of changing the number of \textit{MAN} layers}
\label{amus_layers}
    \begin{tabular}{|c|c|c|}
    \hline
     \textit{Model} & MELD &  IEMOCAP\\\hline
        1-layer & $65.37$ & $68.14$\\\hline
3-layer & $69.26$ & $70.98$\\\hline
5-layer & $71.03$ & $72.57$\\ \hline
\textit{AMUSE} (4 layers) &  $\mathbf {71.32}$ & $\mathbf {73.91}$\\\hline
    \end{tabular}
     \vspace{-5mm}
\end{table} 
\section{Conclusion}
We present \textit{AMuSE} with a \textit{Multimodal Attention Network}, which enables effective knowledge sharing from multiple interactive mode-specific branches to facilitate robust decision-making. 
 Following a multi-loss learning framework, the proposed \textit{Adaptive Fusion} allows  \textit{AMuSE} model to learn the relative contributions of each mode in an effort to learn both category-specific discriminative details and instance-specific contrast-enhanced discriminative cross-modal correspondence within the learned multimodal descriptor. As evident from the experiments,  \textit{AMuSE} delivers a significantly improved performance compared to the baselines. Furthermore, the \textit{Interactive Explainaibility Visualization} also guides the user and produces appropriate mode-wise reasoning for its classification. 
 %Thus, the work exhibits the potential for transformative multi-disciplinary technological research in the areas of community-aware behaviour modelling which are both accurate and explainable.
\vspace{-2mm}
\section*{Acknowledgment}
Computational support is provided by the Center for Computational Research at the University at Buffalo.

%bridges the inter-modal heterogeneity gaps by attending to the mid-level responses of the mode-specific feature representation modules at the hierarchical details. 

\vspace{-2mm}
\bibliographystyle{plain}
\bibliography{references}

\end{document}